\documentclass{llncs}
\usepackage[utf8]{inputenc}
\usepackage[T1]{fontenc}
\usepackage{lmodern}
\usepackage[english]{babel}
\usepackage[babel=true,final]{microtype}
\usepackage{array}
\usepackage{stmaryrd,amsmath,amssymb,mathtools}
\usepackage{braket}
\usepackage{pgfplots}
\usepackage{xfrac}
\usepackage{enumitem}
\usepackage{siunitx}
\usepackage{etoolbox}
\usepackage{bigfoot}
\usepackage{tikz}
\usepackage{algorithm}
\usepackage{algpseudocode}
\usepackage[normalem]{ulem}
\usepackage{pdflscape}
\usepackage{tikz}
\usepackage{qtree}
\usetikzlibrary{positioning}
\usetikzlibrary{arrows,shapes,positioning,calc,patterns,spy,decorations.pathmorphing,decorations}

\DeclareNewFootnote[para]{default}

\newtoggle{longversion}

\newcommand{\angl}[1]{{\langle#1\rangle}}
\newcommand{\sem}[1]{\llbracket#1\rrbracket}

\newcommand{\N}{\mathbb{N}}

\newcommand{\T}{\mathcal{T}}

\renewcommand{\gets}{\coloneqq}
\DeclareMathAlphabet{\kw}{\encodingdefault}{\sfdefault}{bx}{n}
\newcommand{\ignore}[1]{}

\newcommand{\dom}{{\sf dom}}

\newcommand{\equi}{\cong}
\newcommand{\lub}{\bigsqcup}

\newcommand{\prefO}{{\sf pref}_o}
\newcommand{\DeltaO}{\Delta_{out}} 
\newcommand{\DeltaI}{\Delta_{in}} 
\newcommand{\EXP}{\text{EXP}}

\newcommand{\true}{\mathrm{true}}

\pgfplotsset{compat=1.5}

\begin{document}
\pagestyle{plain}
\title{Deciding Equivalence of Separated Non-Nested Attribute Systems in Polynomial Time}
\author{Helmut Seidl\inst{1} \and Raphaela Palenta\inst{1} \and Sebastian Maneth\inst{2}}
\institute{%
	Fakult\"at f\"ur Informatik,
	TU M\"unchen, Germany,\\
	$\{${\tt seidl,palenta}$\}${\tt @in.tum.de}
\and FB3 - Informatik, Universit{\"a}t Bremen, Germany,\\
{\tt maneth@uni-bremen.de}
}
\maketitle

\begin{abstract}
In 1982, Courcelle and Franchi-Zannettacci 
showed that the equivalence problem of separated non-nested attribute systems
can be reduced to the equivalence problem of total deterministic separated basic macro tree transducers. 
They also gave a procedure for deciding equivalence of transducer in the latter class.
Here, we reconsider this equivalence problem.
We present a new alternative decision procedure and prove that 
it runs in polynomial time. 
We also consider extensions of this result to partial transducers and 
to the case where parameters of transducers accumulate strings instead of trees.
\end{abstract}

\section{Introduction}
Attribute grammars are a well-established formalism for realizing 
computations on syntax 
trees~\cite{Knuth1968,DBLP:journals/mst/Knuth71}, and implementations are available for various
programming languages, see, e.g.~\cite{Silver2010,Kiama2013,JavaRAG2015}.
A fundamental question for any such specification formalism is whether two 
specifications are semantically equivalent.
As a particular case, attribute grammars have been considered which compute uninterpreted trees.
Such devices that translate input trees (viz. the parse trees of a context-free grammar) into
output trees, have also been studied under the name ``attributed tree transducer''~\cite{DBLP:journals/actaC/Fulop81}
(see also~\cite{DBLP:series/eatcs/FulopV98}).
In 1982, Courcelle and Franchi-Zannettacci showed that the equivalence problem for (strongly noncircular)
attribute systems reduces to the equivalence problem for primitive recursive schemes with 
parameters~\cite{Courcelle1982,DBLP:journals/tcs/CourcelleF82}; 
the latter model is also known under the name \emph{macro tree transducer}~\cite{DBLP:journals/jcss/EngelfrietV85}.
Whether or not equivalence of attributed tree transducers (ATTs) or of (deterministic) macro tree transducers (MTTs)
is decidable remain two intriguing (and very difficult) open problems.

For several subclasses of ATTs it has been proven that equivalence is decidable.
The most general and very recent result that covers almost all other known ones about
deterministic tree transducers
is that ``deterministic top-down tree-to-string transducers''
have decidable equivalence~\cite{DBLP:journals/jacm/SeidlMK18}. 
Notice that the complexity of this problem remains unknown (the decidability is proved
via two semi-decision procedures).
The only result concerning deterministic tree transducers that we are aware of and that is \emph{not} 
covered by this general result, is the one by 
Courcelle and Franchi-Zannettacci about decidability of equivalence of
``separated non-nested'' ATTs (which they reduce to the same problem for
``separated non-nested'' MTTs). 
However, in their paper no statement is given concerning the complexity of the problem.
In this paper we close this gap and 
study the complexity of deciding equivalence of
separated non-nested MTTs.
To do so we propose a new approach that we feel is
simpler and easier to understand than the one of~\cite{Courcelle1982,DBLP:journals/tcs/CourcelleF82}.
Using our approach we can prove that the
problem can be solved in polynomial time.

\begin{figure}\label{fig:M_tern}
\vspace*{-8mm}
\centering
\Tree [.$g$ [.$f$ [.$f$ [.$f$ $2$ $1$ ] $0$ ] $1$ ] [.$f$ $0$ $2$ ] ]
\Tree [.$+$ [.$+$ [.$*$ $1$ [.EXP $3$ $z$ ] ] [.$+$ [.$*$ $0$ [.EXP $3$ [.$s$ $z$ ] ] ] 
[.$+$
[.$*$ $1$ [.EXP $3$ [.$s$ [.$s$ $z$ ] ] ] ]
[.$*$ $2$ [.EXP $3$ [.$s$ [.$s$ [.$s$ $z$ ] ] ] ] ]
]
] ]
[.$+$ [.$*$ $0$ [.EXP $3$ [.$p$ $z$ ] ] ] [.$*$ $2$ [.EXP $3$ [.$p$ [.$p$ $z$ ] ] ] ] ]
]
\caption{Input tree for 2101.01 (in ternary) and
corresponding output tree of $M_{\text{tern}}$.}
\end{figure}

In a separated non-nested attribute system,
distinct sets of operators are used for the construction of 
inherited and synthesized attributes, respectively, and inherited attributes may depend on
inherited attributes only.
Courcelle and Franchi-Zannettacci's algorithm first translates separated non-nested attribute grammars into separated total deterministic 
non-nested macro tree transducers. In the sequel we will use the more established term
\emph{basic} macro-tree transducers instead of non-nested MTTs.
Here, a macro tree transducer is called \emph{separated} if the alphabets used for the construction of parameter values
and outside of parameter positions are disjoint. 
And the MTT is \emph{basic} if there is no nesting of state calls, i.e., there are no
state calls inside of parameter positions. 
Let us consider an example. We want to translate ternary numbers into expressions over
$+$, $*$, $\EXP$, plus the constants $0$, $1$, and $2$. 
Additionally, operators $s$, $p$, and $z$ are used to represent
integers in unary.
The ternary numbers are parsed into particular binary trees; e.g.,
the left of Figure~\ref{fig:M_tern} shows the binary tree for the number $2101.02$. 
This tree is translated by our MTT into the tree
in the right of Figure~\ref{fig:M_tern} 
(which indeed evaluates to $64.\overline{2}$ in decimal).
The rules of our transducer $M_{\text{tern}}$ are shown in Figure~\ref{fig:rules}.
\begin{figure}[htb]
\[
\begin{array}{lcl}
q_0(g(x_1,x_2))
&\  \to \ \quad&
+(q(x_1,z), q'(x_2,p(z)))\\
q(f(x_1,x_2),y)
&\  \to \ \quad&
+(r(x_2,y), q(x_1,s(y)))\\
q'(f(x_1,x_2),y)
&\ \to \ \quad&
+(r(x_1,y), q'(x_2,p(y)))\\
\phi(i,y)
&\ \to \ \quad&
*(i,\EXP(3,y))\quad\text{for }i\in\{0,1,2\}, \phi\in\{q,q',r\}
\end{array}
\]
\caption{Rules of the transducer $M_{\text{tern}}$.}\label{fig:rules}
\end{figure}
The example is similar to the one used by Knuth~\cite{Knuth1968}
in order to introduce attribute grammars.
The transducer is indeed basic and separated: the operators 
$p$, $s$, and $z$ are only used in parameter positions.

Our polynomial time decision procedure works in two phases:
first, the transducer is converted into an ``earliest'' normal form.
In this form, output symbols that are not produced within parameter positions
are produced as early as possible. In particular it means that the root output
symbols of the right-hand sides of rules for one state must differ.
For instance, our transducer $M_{\text{tern}}$ is \emph{not} earliest,
because all three $r$-rules produce the same output root symbol ($*$).
Intuitively, this symbol should be produced earlier, e.g.,
at the place when the state $r$ is called.
The earliest form is a common technique used for
normal forms and equivalence testing of different kinds of tree transducers~\cite{DBLP:journals/jcss/EngelfrietMS09,DBLP:journals/ijfcs/FrieseSM11,Laurence2011}.
We show that equivalent states of a transducer in this earliest form produce their state-output 
exactly in the same way. This means especially that the output of parameters is 
produced in the same places. It is therefore left
to check, in the second phase, that also these parameter outputs are equivalent. 
To this end, we build an equivalence relation on states of earliest transducers
that combines the two equivalence tests described before.
Technically speaking, the equivalence relation is tested 
by constructing sets of Herbrand equalities.
From these equalities, a fixed point algorithm can, after polynomially many
iterations, produce a stable set of equalities.

An abridged version of this paper will be published within FoSSaCS 2019.

\section{Separated Basic Macro Tree Transducers}\label{s:basics}

Let $\Sigma$ be a ranked alphabet, i.e., every symbol of the finite set 
$\Sigma$ has associated with it a fixed rank $k \in \N$. 
Generally, we assume that the input alphabet $\Sigma$ is \emph{non-trivial}, i.e., 
$\Sigma$ has cardinality
at least 2, and contains at least one symbol of rank $0$ and at least
one symbol of rank $>0$.
The set $\T_\Sigma$ is the set of all (finite, ordered, rooted) trees over the alphabet $\Sigma$.
We denote a tree as a string over $\Sigma$ and parenthesis and commas, i.e.,
$f(a,f(a,b))$ is a tree over $\Sigma$, where $f$ is of rank $2$ and $a,b$ are of rank zero.
We use Dewey dotted decimal notation to refer to a node of a tree:
The root node is denoted $\varepsilon$, and for a node $u$, its $i$-th child
is denoted by $u.i$. For instance, in the tree $f(a,f(a,b))$ the $b$-node is at
position $2.2$.
A \emph{pattern} (or $k$-pattern) (over $\Delta$) is a tree 
$p\in\T_{\Delta\cup\{\top\}}$ over a ranked alphabet $\Delta$ and a disjoint symbol $\top$
(with exactly $k$ occurrences of the symbol $\top$).
The  occurrences of the dedicated symbol $\top$ serve as place holders 
for other patterns.
Assume that $p$ is a $k$-pattern and that $p_1,\dots,p_k$ are patterns;
then $p[p_1,\ldots,p_k]$ denotes 
the pattern obtained from $p$ by replacing, for $i=1,\ldots,k$, the $i$-th occurrence 
(from left-to-right) of
$\top$ by the pattern $p_i$.

A \emph{macro tree transducer} (\emph{MTT}) $M$ is a tuple
$(Q,\Sigma,\Delta,\delta)$ where $Q$ is a ranked alphabet of states,
$\Sigma$ and $\Delta$ are the ranked input and output alphabets, respectively,
and $\delta$ is a finite set of rules of the form:
\begin{eqnarray}
q(f(x_1,\ldots,x_k),y_1,\ldots,y_l)\to T	\label{e:rule}
\end{eqnarray}
where $q\in Q$ is a state of rank $l+1$, $l\geq 0$, $f\in\Sigma$ is an input symbol of rank $k\geq 0$,
$x_1,\ldots,x_k$ and $y_1,\ldots,y_l$ are the formal input and output parameters, respectively,
and $T$ is a tree built up according to the following grammar:
\[
\begin{array}{lll}
T	&{::=}&	a(T_1,\ldots,T_m)
		\mid	q'(x_i,T_1,\ldots,T_n)
		\mid 	y_j
\end{array}
\]
for output symbols $a\in\Delta$ of rank $m\geq 0$ and 
states $q'\in Q$ of rank $n+1$, 
input parameter $x_i$ with $1\leq i\leq k$, and 
output parameter $y_j$ with $1\leq j\leq l$.
For simplicity, we assume that all states $q$ have the same number $l$ of parameters.
Our definition of an MTT does not contain an initial state. We therefore consider an MTT always
together with an axiom $A = p[q_1(x_1,\underline{T_1}),\ldots,q_m(x_1,\underline{T_m})]$ where
$\underline{T_1},\ldots,\underline{T_m} \in \T_\Delta^l$ are vectors of output trees (of length $l$ each).
Sometimes we only use an MTT $M$ without explicitly mentioning an axiom $A$,
then some $A$ is assumed implicitly.
Intuitively, the state $q$ of an MTT corresponds to a function in a functional
language which is defined through pattern matching over its first argument,
and which constructs tree output using tree top-concatenation only;
the second to $(l+1)$-th arguments of state $q$ are its 
accumulating output parameters. 
The output produced by a state
for a given input tree is determined by the right-hand side $T$ 
of a rule of the transducer which
matches the root symbol $f$ of the current input tree.  
This right-hand side is built up from accumulating output parameters and
calls to states for subtrees of the input and applications of output symbols from $\Delta$.
In general MTTs are nondeterministic and only partially defined.
Here, however, we concentrate on total deterministic transducers.
The MTT $M$ is \emph{deterministic}, if for every $(q,f)\in Q\times\Sigma$ there is
at most one rule of the form~\eqref{e:rule}. The MTT $M$ is \emph{total},
if for every $(q,f)\in Q\times\Sigma$ there is
at least one rule of the form~\eqref{e:rule}. 
For total deterministic transducers,
the semantics of a state $q\in Q$ with the rule $q(f(x_1,\ldots,x_k),y_1,\ldots,y_l)\to T$
can be considered as a function
\[
\sem{q}:\T_\Sigma\times\T_\Delta^l\to \T_{\Delta}
\]
which inductively is defined by:
\begin{eqnarray*}
\sem{q}(f(t_1,\ldots,t_k),\underline S) &=&
	\sem{T}\,(t_1,\ldots, t_k)\,\underline S\\
\text{where}	\nonumber	\\
\sem{a(T_1,\ldots,T_m)}\,\underline{t}\,\underline S	&=&	a(\sem{T_1}\,\underline{t}\,\underline S,\ldots,\sem{T_m}\,\underline{t}\,\underline S)	\\
\sem{y_j}\,\underline{t}\,\underline S			&=&	S_j	\\
\sem{q'(x_i,T_1,\ldots,T_l)}\,\underline{t}\,\underline S	&=&	
		\sem{q'}(t_i,\sem{T_1}\,\underline{t}\,\underline S,\ldots,\sem{T_l}\,\underline{t}\,\underline S)	
\end{eqnarray*}
where $\underline S = (S_1,\ldots,S_l)\in\T_\Delta^l$ is a vector of output trees.
The semantics of a pair $(M,A)$ with MTT $M$ and axiom $A = p[q_1(x_1,\underline{T_1}),\ldots,q_m(x_1,\underline{T_m})]$
is defined by $\sem{(M,A)}(t) = p[\sem{q_1}(t,\underline{T_1}),\ldots,\sem{q_m}(t,\underline{T_m})]$.
Two pairs $(M_1,A_1)$, $(M_2,A_2)$ consisting of MTTs $M_1$, $M_2$ and corresponding axioms $A_1$, $A_2$
are \emph{equivalent}, $(M_1,A_1) \equiv (M_2,A_2)$, iff for all input trees $t\in\T_\Sigma$,
and parameter values $\underline{T}\in\T_{\DeltaI}^l$,
$\sem{(M_1,A_1)}(t,\underline{T})=\sem{(M_2,A_2)}(t,\underline{T})$.

The MTT $M$ is \emph{basic}, if each argument tree $T_j$ of a subtree $q'(x_i,T_1,\ldots,T_n)$ 
of right-hand sides $T$ of rules \eqref{e:rule} may not contain further occurrences of states, i.e.,
is in $\T_{\Delta\cup Y}$.
%
%
%
The MTT $M$ is \emph{separated basic}, if $M$ is basic, and 
$\Delta$ is the disjoint union of ranked alphabets
$\DeltaO$ and $\DeltaI$ so that the argument trees $T_j$ of subtrees $q'(x_i,T_1,\ldots,T_n)$
are in $\T_{\DeltaI\cup Y}$, while the output symbols $a$ outside of such subtrees are
from $\DeltaO$. The same must hold for the axiom. Thus, letters directly produced by a state call are in $\DeltaO$ while letters produced in the parameters are in $\DeltaI$.
The MTT $M_{\text{tern}}$ from the Introduction   
is separated basic with $\DeltaO=\{0, 1, 2, 3, *, +, \text{EXP}\}$ and $\DeltaI=\{p,s,z\}$.

As separated basic MTTs are in the focus of our interests, we make the grammar for their right-hand
side trees $T$ explicit:
\[
\begin{array}{lll}
T	&{::=}&	a(T_1,\ldots,T_m)
		\mid 	y_j
		\mid	q'(x_i,T'_1,\ldots,T'_n)	\\
T'	&{::=}&		b(T'_1,\ldots,T'_{m'})
		\mid	y_j
\end{array}
\]
where $a\in\DeltaO$, $q'\in Q$, $b\in\DeltaI$ of ranks $m,n+1$ and $m'$, respectively,
and $p$ is an $n$-pattern over $\Delta$.
For separated basic MTTs only axioms $A=p[q_1(x_1,\underline{T_1}),\ldots,q_m(x_1,\underline{T_m})]$ with $T_1,\ldots,T_m \in \T_{\DeltaI}^l$ are considered.

Note that equivalence of nondeterministic transducers is undecidable (even already for
very small subclasses of transductions~\cite{DBLP:journals/jacm/Griffiths68}). 
Therefore, we assume for the rest of the paper that all 
MTTs are deterministic and separated basic.
We will also assume that all MTTs are total, with the exception of Section~\ref{sec:app}
where we also consider partial MTTs.
\begin{example}\label{ex:binary}
We reconsider the example from the Introduction and adjust it to our formal definition. 
The transducer was given without an axiom (but with a
tacitly assumed ``start state'' $q_0$).
Let us now remove the state $q_0$ and
add the axiom $A=q(x_1,z)$. 
The new $q$ rule for $g$ is:
\[
q(g(x_1,x_2),y)\to +(q(x_1,y),q'(x_2,p(y))).
\]
To make the transducer total, 
we add for state $q'$ the rule
\[
q'(g(x_1,x_2),y) \to +(*(0,\text{EXP}(3,y)),*(0,\text{EXP}(3,y))).
\]
For state $r$ we add rules $q(\alpha(x_1,x_2),y) \to *(0,\text{EXP}(3,y))$ with $\alpha = f,g$.
The MTT is separated basic with $\DeltaO=\{0, 1, 2, 3, *, +, \text{EXP}\}$ and $\DeltaI=\{p,s,z\}$.
\qed
\end{example}

We restricted ourselves to \emph{total} separated basic MTTs. However, we would
like to be able to decide equivalence
for \emph{partial} transducers as well.
For this reason we define now top-down tree automata, and will then decide equivalence 
of MTTs relative to some given DTA $D$.
A \emph{deterministic top-down tree automaton} (\emph{DTA}) $D$ is a tuple $(B, \Sigma, b_0, \delta_D)$ where
$B$ is a finite set of states, $\Sigma$ is a ranked alphabet of input symbols, $b_0\in B$ is the initial state,
and $\delta_D$ is the partial transition function with rules of the form 
$b(f(x_1,\ldots, x_k)) \to (b_1(x_1),\ldots, b_k(x_k))$,
where $b,b_1,\dots,b_k\in B$ and $f\in\Sigma$ of rank $k$.
W.l.o.g.\ we always assume that all states $b$ of a DTA are productive, i.e., $\dom(b)\neq\emptyset$.
If we consider a MTT $M$ relative to a DTA $D$ we implicitly assume a mapping $\pi: Q \to B$,
that maps each state of $M$ to a state of $D$, then we consider for $q$ only input trees
in $\dom(\pi(q))$.

\section{Top-Down Normalization of Transducers}\label{s:earliest}

In this section we show that each total deterministic basic separated MTT
can be put into an ``earliest'' normal form relative to a fixed DTA $D$.
Intuitively, state output
(in $\DeltaO$) is produced as early as possible for a transducer in the normal
form. It can then be shown that two equivalent transducers in normal form
produce their state output in exactly the same way.

Recall the definition of patterns as trees over $\T_{\Delta\cup\{\top\}}$.
Substitution of $\top$-symbols by other patterns
induces a partial ordering $\sqsubseteq$ over patterns by
$p\sqsubseteq p'$ if and only if $p = p'[p_1,\ldots,p_m]$ for some patterns $p_1,\ldots,p_m$.
W.r.t.\ this ordering, $\top$ is the \emph{largest} element, while all patterns
without occurrences of $\top$ are minimal.
By adding an artificial \emph{least} element $\bot$, the resulting partial ordering is
in fact a \emph{complete lattice}. Let us denote this complete lattice by $\mathcal{P}_\Delta$.

Let $\Delta=\DeltaI\cup \DeltaO$.
For $T\in\T_{\Delta\cup Y}$, we define the \emph{$\DeltaO$-prefix} as the 
pattern $p\in\T_{\DeltaO\cup\{\top\}}$ as follows.
Assume that $T = a(T_1,\ldots,T_m)$.
\begin{itemize}
\item
If $a\in\DeltaO$, then
$p = a(p_1,\ldots,p_m)$ where for $j=1,\ldots,m$, $p_j$ is the $\DeltaO$-prefix of $T_j$.
\item
If $a\in\DeltaI\cup Y$, then $p = \top$.
\end{itemize}
By this definition, each tree $t\in\T_{\Delta\cup Y}$ can be uniquely decomposed into a
$\DeltaO$-prefix $p$ and subtrees $t_1,\ldots,t_m$ whose root symbols all are contained in
$\DeltaI\cup Y$ such that $t = p[t_1,\ldots,t_m]$.

Let $M$ be a total separated basic MTT $M$,
$D$ be a given DTA.
We define the $\DeltaO$-prefix of a state $q$ of $M$ relative to $D$
as the minimal pattern $p\in\T_{\DeltaO\cup\{\top\}}$
so that each tree $\sem{q}(t,\underline T)$, $t\in\dom(\pi(q)),\underline T\in\T_\Delta^l,$ 
is of the form 
$p[T_1,\ldots,T_m]$ for some sequence of subtrees $T_1,\ldots,T_m \in \T_{\Delta}$.
Let us denote this unique pattern $p$ by $\prefO(q)$. 
If $q(f, y_1,\ldots, y_l) \rightarrow T$ is a rule of a separated basic MTT and there is
an input tree $f(t_1,\ldots, t_k) \in \dom(\pi(q))$ then $|\prefO(q)| \leq |T|$.

\begin{lemma}\label{l:prefix}
Let $M$ be a total separated basic MTT and $D$ a given DTA.
Let $t\in\dom(\pi(q))$ be a smallest input tree of a state $q$ of $M$.
The $\DeltaO$-prefix of every state $q$ of $M$ relative to $D$
can be computed in time $\mathcal{O}(|t|\cdot |M|)$.
\end{lemma}
The proof is similar to the one of \cite[Theorem 8]{DBLP:journals/jcss/EngelfrietMS09} 
for top-down tree transducers.
This construction can be carried over as, for the computation of $\DeltaO$-prefixes,
the precise contents of the output parameters $y_j$ can be ignored.
The complete proof can be found in the Appendix.

\begin{example}\label{ex:prefO}
 We compute the $\DeltaO$-prefix of the MTT $M$ from Example \ref{ex:binary}.
 We consider $M$ relative to the trivial DTA $D$ that consists only of one state $b$ with
 $dom(b) = \T_\Sigma$. We therefore omit $D$ in our example.
 We get the following system of in-equations: from the rules of state $r$ we obtain
 $Y_r \sqsubseteq *(i,\text{EXP}(3,\top))$ with $i \in \{0,1,2\}$.
 From the rules of state $q$ we obtain
 $Y_q \sqsubseteq +(Y_q,Y_{q'})$, $Y_q \sqsubseteq +(Y_r,Y_{q})$ and
 $Y_q \sqsubseteq *(i,\text{EXP}(3,\top))$ with $i \in \{0,1,2\}$.
 From the rules of state $q'$ we obtain
 $Y_{q'} \sqsubseteq +(*(0,\text{EXP}(3,\top)),*(0,\text{EXP}(3,\top)))$, $Y_{q'} \sqsubseteq +(Y_r,Y_{q'})$ and
 $Y_{q'} \sqsubseteq *(i,\text{EXP}(3,\top))$ with $i \in \{0,1,2\}$.
 For the fixpoint iteration we initialize $Y_r^{(0)}$, $Y_q^{(0)}$, $Y_{q'}^{(0)}$ with $\bot$ each.
 Then $Y_r^{(1)} = *(\top,\text{EXP}(3,\top)) = Y_r^{(2)}$ and
 $Y_q^{(1)} = \top$, $Y_{q'}^{(1)} = \top$.
 Thus, the fixpoint iteration ends after two rounds with the solution $\prefO(q) = \top$.
\qed
\end{example}

Let $M$ be a separated basic MTT $M$ and $D$ be a given DTA $D$.
$M$ is called $D$-earliest if for every state $q \in Q$ the $\DeltaO$-prefix with respect
to $\pi(q)$ is $\top$.

\begin{lemma}\label{l:earliest}
For every pair $(M,A)$ consisting of a total  separated basic MTT $M$ and axiom $A$
and a given DTA $D$, 
an equivalent pair $(M',A')$
can be constructed so that
$M'$ is a total  separated basic MTT that is $D$-earliest.
Let $t$ be an output tree of $(M,A)$ for a smallest input tree $t' \in\dom(\pi(q))$
where $q$ is the state occurring in $A$.
Then the construction runs in time $\mathcal{O}(|t|\cdot|(M,A)|)$.
\end{lemma}

The construction follows the same line as the one for the earliest form of top-down tree
transducer, cf. \cite[Theorem 11]{DBLP:journals/jcss/EngelfrietMS09}.
The proof can be found in the appendix.
\noindent
Note that for partial separated basic MTTs the size of the $\DeltaO$-prefixes is at most exponential
in the size of the transducer.
However for total transducer that we consider here
the $\DeltaO$-prefixes are linear in the size
of the transducer and can be computed in quadratic time, cf.~\cite{DBLP:journals/jcss/EngelfrietMS09}.
\begin{corollary}\label{c:total}
For $(M,A)$ consisting of a total deterministic separated basic MTT $M$ and axiom $A$
and the trivial DTA $D$ accepting $\T_\Sigma$ an equivalent pair $(M',A')$ can be
constructed in quadratic time such that $M'$ is an $D$-earliest total deterministic separated
basic MTT.
\end{corollary}

\begin{example}
We construct an equivalent earliest MTT $M'$ for the transducer from Example \ref{ex:binary}.
In Example \ref{ex:prefO} we already computed the $\DeltaO$-prefixes of states $q, q', r$;
$\prefO(q) = \top$, $\prefO(q') = \top$ and $\prefO(r) = *(\top, \text{EXP}(3,\top))$.
As there is only one occurrence of symbol $\top$ in the $\DeltaO$-prefixes of $q$ and $q'$
we call states $\angl{q,1}$ and $\angl{q',1}$ by $q$ and $q'$, respectively.
Hence, a corresponding earliest transducer has
axiom $A=q(x_1,z)$.
The rules of $q$ and $q'$ for input symbol $g$ do not change.
For input symbol $f$ we obtain
\[
\begin{array}{rclr}
q(f(x_1, x_2), y) &\to& +(*(r(x_2,y),\text{EXP}(3,y)), q(x1,s(y))) &\text{ and} \\
q'(f(x_1,x_2), y) &\to& +(*(r(x_1,y),\text{EXP}(3,y),q'(x_2,p(y))).
\end{array}
\]

As there is only one occurrence of symbol $\top$ related to a recursive call in $\prefO(r)$
we call $\angl{r,1}$ by $r$.
For state $r$ we obtain new rules $r(\alpha(x_1,x_2),y) \to 0$ with
$\alpha \in \{f,g\}$ and
$r(i,y) \to i$ with $i \in \{0,1,2\}$.
\qed
\end{example}

We define a family of equivalence relation by induction, 
$\equi_b\ \subseteq ((Q,\T_{\DeltaI}^k) \cup \T_{\DeltaI}) \times ((Q,\T_{\DeltaI}^k) \cup \T_{\DeltaI})$
with $b$ a state of a given DTA is the intersection of the equivalence relations $\equi^{(h)}_b$, i.e., $X \equi_b Z$ if and only if for all $h \geq 0$,
$X \equi^{(h)}_b Z$.
We let $(q,\underline{T}) \equi^{(h+1)}_b (q',\underline{T'})$ if
for all $f \in \dom(b)$ with $b(f(x_1,\ldots,x_k) \to (b_1,\ldots, b_k)$,
there is a pattern $p$ such that
$q(f(x_1,\ldots, x_k),\underline{y}) \rightarrow p[t_1,\ldots, t_m]$ and
$q'(f(x_1,\ldots, x_k),\underline{y'}) \rightarrow p[t'_1,\ldots,t'_m]$
 with
\begin{itemize}
  \item if $t_i$ and $t'_i$ are both recursive calls to the same subtree, i.e., 
  $t_i = q_i(x_{j_i},\underline{T_i})$, $t'_i = q'_i(x_{j'_i},\underline{T'_i})$ and $j_i = j'_i$, then
  $(q_i,\underline{T_i})[\underline{T}/\underline{y}] \equi^h_{b_{j_i}} (q'_i,\underline{T'_i})[\underline{T'}/\underline{y'}]$
  \item if $t_i$ and $t'_i$ are both recursive calls but on different subtrees, i.e., 
  $t_i = q_i(x_{j_i},\underline{T_i})$, $t'_i = q'_i(x_{j'_i},\underline{T'_i})$ and $j_i \neq j'_i$, then
  $\hat{t} := \sem{q_i}(s,\underline{T_i})[\underline{T}/\underline{y}] = \sem{q'_i}(s,\underline{T'_i})[\underline{T}/\underline{y}]$ for some $s \in \Sigma^{(0)}$ and 
  $(q_i,\underline{T_i})[\underline{T}/\underline{y}] \equi^{(h)}_{b_{j_i}} \hat{t} \equi^{(h)}_{b_{j'_i}} (q'_i,\underline{T'_i})[\underline{T}/\underline{y}]$ 
  \item if $t_i$ and $t'_i$ are both parameter calls, i.e., $t_i = y_{j_i}$ and $t'_{i} = y'_{j'_i}$, then $T_{j_i} = T'_{j'_i}$
  \item if $t_i$ is a parameter call and $t'_i$ a recursive call, i.e., $t_i = y_{j_i}$ and $t'_i = q'_i(x_{j'_i},\underline{T'_i})$,
  then $T_{j_i} \equi^{(h)}_{b_{j'_i}} (q'_i,\underline{T'_i})[\underline{T'}/\underline{y'}]$
  \item (symmetric to the latter case) if $t_i$ is a recursive call and $t'_i$ a parameter call, i.e.,
  $t_i = q_i(x_{j_i},\underline{T_i})$ and $t'_i = y'_{j'_i}$, then 
  $(q_i,\underline{T_i})[\underline{T}/\underline{y}] \equi^{(h)}_{b_{j_i}} T'_{j'_i}$.
\end{itemize}
We let $T \equi^{(h+1)}_b (q',\underline{T'})$ if for all $f \in \dom(b)$ with $r(f(x_1,\ldots, x_k))\to (b_1,\ldots, b_k)$,
$q'(f(\underline{x}),\underline{y}) \rightarrow t'$,
\begin{itemize}
  \item if $t' = y_j$ then $T = T'_j$
  \item if $t' = q'_1(x_i,\underline{T'_1})$ then $T \equi^{(h)}_{b_i} (q'_1,\underline{T'_1})[\underline{T'}/\underline{y'}]$.
\end{itemize}
Intuitively, $(q,\underline{T}) \equi^{h}_b (q',\underline{T'})$ if
for all input trees $t \in \dom(b)$ of height $h$,
$\sem{q}(t,\underline{T}) = \sem{q'}(t,\underline{T'})$.
Then $(q,\underline{T}) \equi_b (q',\underline{T'})$ if
for \emph{all} input trees $t \in \dom(b)$ (independent of the height),
$\sem{q}(t,\underline{T}) = \sem{q'}(t,\underline{T'})$.

\begin{theorem}
For a given DTA $D$ with initial state $b$, 
let $M, M'$ be $D$-earliest 
total deterministic separated basic MTTs with axioms $A$ and $A'$, respectively.
Then $(M,A)$ is equivalent to $(M',A')$ relative to $D$, iff
there is a pattern $p$ such that $A=p[q_1(x_1,\underline{T_1}),\ldots, q_m(x_1,\underline{T_m})]$, and 
$A'=p[q'_1(x_1,\underline{T'_1}),\ldots, q'_m(x_1,\underline{T'_m})]$ and for $j=1,\ldots,m$,
$(q_j,\underline{T_j}) \equi_b (q'_j,\underline{T'_j})$, i.e., $q_j$ and $q'_j$ are equivalent 
on the values of output parameters $\underline T_j$ and $\underline T'_j$.
\end{theorem}
\begin{proof}
Let $\Delta$ be the output alphabet of $M$ and $M'$.
Assume that $(M,A) \equi_b (M',A')$. As $M$ and $M'$ are earliest, the $\DeltaO$-prefix of
$\sem{(M,A)}(t)$ and $\sem{(M',A')}(t)$, for $t \in \dom(b)$ is the same pattern $p$
and therefore $A=p[q_1(x_1,\underline{T_1}),\ldots, q_m(x_1,\underline{T_m})]$ and
$A'=p[q'_1(x_1,\underline{T'_1}),\ldots, q'_m(x_1,\underline{T'_m})]$.
To show that $(q_i,\underline{T_i}) \equi_{b} (q'_i,\underline{T'_i})$ let $u_i$ be the position of the $i$-th $\top$-node
in the pattern $p$. For some $t\in \dom(b)$ and $\underline{T}\in\T_{\DeltaI}$ let $t_i$ and $t'_i$
be the subtree of $\sem{(M,A)}(t,\underline{T})$ and $\sem{(M',A')}(t,\underline{T})$, respectively.
Then $t_i = t'_i$ and therefore $(q_i,\underline{T_i}) \equi_b (q'_i,\underline{T'_i})$.

Now, assume that the axioms $A=p[q_1(x_1,\underline{T_1}),\ldots, q_m(x_1,\underline{T_m})]$ and
$A'=p[q'_1(x_1,\underline{T'_1}),\ldots, q'_m(x_1,\underline{T'_m})]$ consist of the same pattern $p$
and for $i=1,\ldots,m$, $(q_i,\underline{T_i}) \equi_b (q'_i,\underline{T'_i})$.
Let $t\in \dom(b)$ be an input tree then
\[
\begin{array}{lll}
\sem{(M,A)}(t) &=& p[\sem{q_1}(t,\underline{T_1}),\ldots,\sem{q_m}(t,\underline{T_m})]\\
			    &=& p[\sem{q'_1}(t,\underline{T'_1}),\ldots,\sem{q'_m}(t,\underline{T'_m})] \\
			    &=& \sem{(M',A')}(t).
\end{array}
\]
\end{proof}

\section{Polynomial Time}\label{s:poly}

In this section we prove the main result of this paper, namely,
that for each fixed DTA $D$,
equivalence of total deterministic basic separated MTTs 
(relative to $D$) can be decided in polynomial time. 
This is achieved by taking as input
two $D$-earliest such transducers, and then collecting conditions on the parameters
of pairs of states of the respective transducers for their produced outputs to be equal.

\begin{example}\label{e:eq}
Consider a DTA $D$ with a single state only which accepts all inputs, and states $q,q'$ with
\[
\begin{array}{lll@{\qquad}lll}
q(a,y_1,y_2)	&\to&	g(y_1)	&
q'(a,y'_1,y'_2)	&\to&	g(y'_2)
\end{array}
\]
Then $q$ and $q'$ can only produce identical outputs for the input $a$ (in $\dom(b)$)
if parameter $y_2'$ of $q'$ contains the same output tree as
parameter $y_1$ of $q$. This precondition can be formalized by the equality
$y_2'\doteq y_1$.
Note that in order to distinguish the output parameters of $q'$ from those of $q$
we have used primed copies $y'_i$ for $q'$.
\qed
\end{example}
It turns out that \emph{conjunctions} of equalities such as in example \ref{e:eq}
are sufficient for proving equivalence of states.
\noindent 
For states $q, q'$ of total separated basic MTTs $M, M'$, respectively,
that are both $D$-earliest for some
fixed DTA $D$, $h \geq 0$ and some fresh variable $z$, we define
\[\begin{array}{lclr}
 \displaystyle \Psi_{b,q}^{(h)}(z) = \bigwedge_{b(f\underline{x}) \to (b_1,\ldots, b_k)} &\displaystyle\bigwedge_{q(f\underline{x},\underline{y})\rightarrow y_j}& \displaystyle(z \doteq y_j) &\wedge\\
					   &\displaystyle\bigwedge_{q(f\underline{x},\underline{y})\rightarrow \hat{q}(x_i,\underline{T})}& \displaystyle\Psi_{b_i,\hat{q}}^{(h-1)}(z)[\underline{T}/\underline{y}] &\wedge\\
					   & \displaystyle\bigwedge_{q(f\underline{x},\underline{y})\rightarrow p[\ldots] \atop p \neq \top}&\displaystyle \bot
\end{array}\]
where $\bot$ is the boolean value \emph{false}.
We denote the output parameters in $\Psi_{b,q}^{(h)}(z)$ by $\underline{y}$,
we define $\Psi_{b,q'}^{\prime(h)}(z)$ in the same lines
as $\Psi_{b,q}^{(h)}(z)$ but using $\underline{y'}$ for the output parameters. 
To substitute the output parameters with 
trees $\underline{T}$, $\underline{T'}$, we therefore use 
$\Psi_{b,q}^{(h)}(z)[\underline{T}/\underline{y}]$ and
$\Psi_{b,q'}^{\prime(h)}(z)[\underline{T'}/\underline{y'}]$.
Assuming that $q$ is a state of the $D$-earliest separated basic MTT $M$ then 
$\Psi_{b,q}^{(h)}(z)$ is true for all 
ground parameter values $\underline s$ and some $T\in\T_{\Delta\cup Y}$ if 
$\sem{q}(t,\underline s) = T[\underline s/\underline y]$ 
for all input trees $t\in\dom(b)$ of height at most $h$.
Note that, since $M$ is $D$-earliest, $T$ is necessarily in $\T_{\DeltaI\cup Y}$.
%
%
W.l.o.g., we assume that every state $b$ of $D$ is productive, i.e., $\dom(b)\neq\emptyset$.
For each state $b$ of $D$, we therefore may choose some input tree $t_b\in\dom(b)$ of 
minimal depth.
We define $s_{b,q}$ to be the output of $q$ for a minimal input tree $t_r \in \dom(b)$
and parameter values $\underline{y}$ --- when considering formal output parameters
as output symbols in $\DeltaI$,
i.e., $s_{b,q} = \sem{q}(t_r,\underline{y})$.
\begin{example}
We consider again the trivial DTA $D$ with only one state $b$ that accepts all $t \in \T_\Sigma$.
Thus, we may choose $t_b = a$.
For a state $q$ with the following two rules 
$q(a,y_1,y_2) \rightarrow y_1$ and 
$q(f(x),y_1,y_2) \rightarrow q(x,h(y_2),b)$, we have $s_{b,q} = y_1$. 
Moreover, we obtain
\begin{eqnarray*}
\Psi_{b,q}^{(0)}(z) &=& z \doteq y_1 \\
\Psi_{b,q}^{(1)}(z) &=& (z \doteq y_1) \wedge (z \doteq h(y_2)) \\
\Psi_{b,q}^{(2)}(z) &=& (z \doteq y_1) \wedge (z \doteq h(y_2)) \wedge (z \doteq h(b))\\
	      &\equiv& (y_2 \doteq b) \wedge (y_1 \doteq h(b)) \wedge (z \doteq h(b)) \\
\Psi_{b,q}^{(3)}(z) &=& (z \doteq y_1) \wedge (b \doteq b) \wedge (h(y_2) \doteq h(b)) \wedge (z\doteq h(b))\\
	      &\equiv& (y_2 \doteq b) \wedge (y_1 \doteq h(b)) \wedge (z \doteq h(b))
\end{eqnarray*}
We observe that $\Psi_{b,q}^{(2)}(z) = \Psi_{b,q}^{(3)}(z)$ and therefore for every $h\geq 2$, $\Psi_{b,q}^{(h)}(z) = \Psi_{b,q}^{(3)}(z)$.
\qed
\end{example}

\noindent
According to our equivalence relation $\equi_b$,
$b$ state of the DTA $D$,
we define for states $q,q'$
of $D$-earliest total deterministic separated basic MTTs $M,M'$, 
and $h \geq 0$, the conjunction $\Phi_{b,(q,q')}^{(h)}$ by
\[
\begin{array}{llclr}
 &\displaystyle \bigwedge_{b(f\underline{x}) \to (b_1,\ldots,b_k) \atop{q(f\underline{x},\underline{y}) \rightarrow p[\underline{t}] \atop q'(f\underline{x},\underline{y'})\rightarrow p[\underline{t'}]}} \Big( &\displaystyle \bigwedge_{t_i = y_{j_i}, \atop t'_i = y'_{j'_i}}& \displaystyle(y_{j_i} \doteq y'_{j'_i}) &\wedge \\
		   &   &\displaystyle \bigwedge_{t_i = y_{j_i}, \atop t'_i = q'_i(x_{j'_i}, \underline{T'})}& \displaystyle \Psi_{b_{j'_i},q'_i}^{\prime(h-1)}(y_{j_i})[\underline{T'}/\underline{y'}] &\wedge  \\
		   &   &\displaystyle \bigwedge_{t'_i = y'_{j'_i}, \atop t_i = q_i(x_{j_i},\underline{T})}& \displaystyle \Psi_{b_{j_i},q_i}^{(h-1)}(y'_{j'_i})[\underline{T}/\underline{y}] &\wedge \\
		   &  &\displaystyle \bigwedge_{t_i = q_i(x_{j_i},\underline{T}), \atop{t'_i = q'_i(x_{j'_i},\underline{T'}) \atop j_i = j'_i}}& \displaystyle \Phi_{b_{j_i}, (q_i,q'_i)}^{(h-1)}[\underline{T}/\underline{y},\underline{T'}/\underline{y'}] &\wedge \\
		   &  &\displaystyle \bigwedge_{t_i = q_i(x_{j_i},\underline{T}), \atop{t'_i = q'_i(x_{j'_i},\underline{T'}) \atop j_i \neq j'_i}}& \displaystyle 
	(\Psi_{b_{j_i},q_i}^{(h-1)}(s_{b,q_i})[\underline{T}/\underline{y}] \wedge \Psi_{b_{j'_i},q'_i}^{\prime(h-1)}(s_{b,q_i}[\underline{T}/\underline y])[\underline{T'}/\underline{y'}])\ \Big) &\wedge \\
		    &\displaystyle \bigwedge_{b(f) \to (b_1,\ldots,b_k)  \atop{p \neq p',q(f\underline{x},\underline{y}) \rightarrow p[\underline{t}] \atop q'(f\underline{x},\underline{q})\rightarrow p'[\underline{t'}]}} & \displaystyle \bot
\end{array}
\]
$\Phi_{b,(q,q')}^{(h)}$ is defined in the same lines as the equivalence relation $\equi^{(h)}_b$. 
$\Phi_{b,(q,q')}^{(h)}$ is true for all values of output parameters $\underline{T}$, $\underline{T'}$ such that
$\sem{q}(t,\underline{T}) = \sem{q'}(t,\underline{T'})$ for $t \in\dom(b)$ of height at most $h$.
By induction on $h\geq 0$, we obtain:

\begin{lemma}\label{l:PhiEquiv}
For a given DTA $D$,
states $q,q'$ of $D$-earliest total separated basic MTTs, 
vectors of trees $\underline{T}, \underline{T'}$ over $\DeltaI$, $b$ a state of $D$.
$s\in\dom(b)$,
and $h\geq 0$ the following two statements hold:
 $$(q,\underline{T}) \equi_b^{(h)} (q',\underline{T'}) \Leftrightarrow \Phi_{b,(q,q')}^{(h)}[\underline{T}/\underline{y},\underline{T'}/\underline{y'}] \equiv \true$$
$$s \equi_b^{(h)} (q',\underline{T'}) \Leftrightarrow \Psi_{b,q'}^{(h)}(t)[\underline{T'}/\underline{y}] \equiv \true$$
\qed
\end{lemma}

\noindent
$\Phi_{b,(q,q')}^{(h)}$ is a conjunction of equations of the form $y_i \doteq y_j$, $y_i \doteq t$ with $t \in \DeltaI$.
Every satisfiable conjunction of equalities is equivalent to a (possible empty) finite
conjunction of equations of the form $y_{i} \doteq t_{ i}$, $t_i\in\T_{\DeltaI\cup Y}$ where the $y_{i}$
are distinct and no equation is of the form $y_j\doteq y_j$. 
We call such conjunctions \emph{reduced}. If we have two inequivalent reduced conjunctions $\phi_1$ and $\phi_2$ with
$\phi_1 \Rightarrow \phi_2$ then $\phi_1$ contains strictly more equations. From that follows that for every
sequence $\phi_0 \Rightarrow \ldots \phi_m$ of pairwise inequivalent reduced conjunctions $\phi_j$ with $k$ variables,
$m \leq k+1$ holds.
This observation is crucial for the termination of the fixpoint iteration we will use to compute $\Phi_{b,(q,q')}^{(h)}$.

For $h\geq 0$ we have:
\begin{align}
 \Psi_{b,q}^{(h)}(z) &\Rightarrow \Psi_{b,q}^{(h-1)}(z) \\
 \Phi_{b,(q,q')}^{(h)} &\Rightarrow \Phi_{b,(q,q')}^{(h-1)} \label{eq:hh1}
\end{align}

As we fixed the number of output parameters to the number $l$, for each pair $(q,q')$ the conjunction $\Phi_{b,(q,q')}^{(h)}$ contains at most $2l$ variables $y_i, y'_i$.  Assuming that the MTTs to which state
$q$ and $q'$ belong have $n$ states each, we conclude that
$\Phi_{b,(q,q')}^{(n^2(2l+1))} \equiv \Phi_{b,(q,q')}^{(n^2(2l+1)+i)}$ and
$\Psi_{b,q}^{(n(l+1))} \equiv \Psi_{b,q}^{(n(l+1)+i)}$ for all $i \geq 0$. 
Thus, we can define $\Phi_{b,(q,q')} := \Phi_{b,(q,q')}^{(n^2(2l+1))}$
and $\Psi_{b,q} := \Psi_{b,q}^{(n(l+1))}$. 
As $(q,\underline{T}) \equi_b (q',\underline{T'})$ iff for all $h \geq 0$,
$(q,\underline{T}) \equi_b^{(h)} (q',\underline{T'})$ holds, observation (\ref{eq:hh1}) 
implies that
$$(q,\underline{T}) \equi_b (q',\underline{T'}) \Leftrightarrow \Phi_{b,(q,q')}[\underline{T}/\underline{y}][\underline{T'}/\underline{y'}] \equiv \true$$
Therefore, we have:

\begin{lemma}\label{l:Phi}
For a DTA $D$,
states $q, q'$ of $D$-earliest separated basic MTTs $M,M'$ and states $b$ of $D$,
the formula $\Phi_{b,(q,q')}$ can be computed in time polynomial in the sizes of $M$ and
$M'$.
%
%
\end{lemma}

\begin{proof}
We successively compute  the conjunctions 
$\Psi_{b,q}^{(h)}(z), \Psi_{b,q'}^{(h)}(z), \Phi_{b,(q,q')}^{(h)}, h\geq 0$, for all states
$b,q,q'$. 
As discussed before, some $h\leq n^2(2l+1)$ exists such that the conjunctions for $h+1$
are equivalent to the corresponding conjunctions for $h$ --- in which case, we terminate.
It remains to prove that the conjunctions for $h$ can be computed from the conjunctions 
for $h-1$ in polynomial time.
For that, it is crucial that we maintain \emph{reduced} conjunctions.
Nonetheless, the \emph{sizes} of occurring right-hand sides of equalities 
may be quite large.
Consider for example the conjunction 
$x_1 \doteq a \wedge x_2 \doteq f(x_1,x_1) \wedge \ldots \wedge x_n \doteq f(x_{n-1},x_{n-1})$. 
The corresponding reduced conjunction is then given by
$x_1 \doteq a\wedge x_2 \doteq f(a,a)\wedge \ldots\wedge x_n \doteq f(f(f(\ldots(f(a,a))\ldots)$
where the sizes of right-hand sides  grow exponentially.
In order to arrive at a polynomial-size representation, we therefore rely on compact representations
where isomorphic subtrees are represented only once.
W.r.t.\ this representation, reduction of a non-reduced conjunction,
implications between reduced conjunctions as well as 
substitution of variables in conjunctions can all be realized in polynomial time.
From that, the assertion of the lemma follows.
\end{proof}

\begin{example}
Let $D$ be a DTA with the following rules $b(f(x)) \to (b)$, $b(g) \to ()$ and
$b(h) \to ()$. 
Let $q$ and $q'$ be states of separated basic MTTs $M$, $M'$, respectively,
that are $D$-earliest
and $\pi$, $\pi'$ be the mappings from the states of $D$ to the states of $M$, $M'$ with
$(b,q)\in \pi$ and $(b,q')\in\pi'$.
\[\begin{array}{rcl}
      q(f(x),y_1,y_2) &\rightarrow& a(q(x,b(y_1,y_1),c(y_2),d))\\
      q(g,y_1,y_2) &\rightarrow& y_1 \\
      q(h,y_1,y_2) &\rightarrow& y_2
    
   \end{array}
\]
\[\begin{array}{rcl}
      q'(f(x),y'_1,y'_2) &\rightarrow& a(q'(x,c(y'_1),b(y'_2,y'_2),d))\\
      q'(g,y'_1,y'_2) &\rightarrow& y'_2 \\
      q'(h,y'_1,y'_2) &\rightarrow& y'_1
    
   \end{array}
\]
\begin{eqnarray*}
 \Phi_{r,(q,q')}^{(0)} &=& (y_1 \doteq y'_2) \wedge (y_2 \doteq y'_1) \\
 \Phi_{r,(q,q')}^{(1)} &=& (y_1 \doteq y'_2) \wedge (y_2 \doteq y'_1) \wedge (b(y_1,y_1) \doteq b(y'_2,y'_2)) \wedge (c(y_2) \doteq c(y'_1)) \\
		  &\equiv& (y_1 \doteq y'_2) \wedge (y_2 \doteq y'_1) = \Phi_{r,(q,q')}^{(0)}
\end{eqnarray*}
\qed
\end{example}

\noindent
In summary, we obtain the main theorem of our paper.

\begin{theorem}\label{t:eqpo}
Let $(M,A)$ and $(M',A')$ be pairs consisting of total deterministic separated basic MTTs $M$, $M'$ 
and corresponding axioms $A$, $A'$ and $D$ a DTA.
Then the equivalence of $(M,A)$ and $(M',A')$ relative to $D$ is decidable.
If $D$ accepts all input trees, equivalence can be decided in polynomial time.
\end{theorem}

\begin{proof}
By Lemma \ref{l:earliest} we build pairs $(M_1,A_1)$ and $(M'_1, A'_1)$ that are equivalent to $(M,A)$ and $(M',A')$
where $M_1$, $M'_1$ are $D$-earliest separated basic MTTs.
If $D$ is trivial the construction is in polynomial time, cf.\ Corollary~\ref{c:total}.
Let the axioms be $A_1 = p[q_1(x_{i_1},\underline{T_1}),\ldots, q_k(x_{i_k},\underline{T_k})]$ and 
$A'_1 = p'[q'_1(x_{i'_1}, \underline{T_1}),\ldots,q'_k(x_{i'_{k'}},\underline{T_{k'}})]$.
According to Lemma \ref{l:PhiEquiv} $(M_1,A_1)$ and $(M'_1, A'_1)$ are equivalent
iff
\begin{itemize}
 \item $p = p'$, $k = k'$ and
 \item for all $j = 1,\ldots, k$, $\Phi_{b,(q_j,q'_j)}[\underline{T_j}/\underline{y},\underline{T_j}/\underline{y'}]$ is
 equivalent to ${\sf true}$.
\end{itemize}
By Lemma~\ref{l:Phi} we can decide the second statements in time polynomial in the sizes of
$M_1$ and $M'_1$.
\end{proof}

\section{Applications} \label{sec:app}


In this section we show several applications of our equivalence result.
First, we consider partial transductions of separated basic MTTs.
To decide the equivalence of partial transductions we need to decide
$a)$ whether the domain of two given MTTs is the same and if so,
$b)$ whether the transductions on this domain are the same.
How the second part of the decision procedure is done was shown in detail in this paper
if the domain is given by a DTA. It therefore remains to discuss how this DTA can be obtained.
It was shown in \cite[Theorem 3.1]{DBLP:journals/mst/Engelfriet77} that the domain of every
top-down tree transducer $T$ can be accepted by some DTA $B_T$ and this automaton can be
constructed from $T$ in exponential time. This construction can easily be extended to basic MTTs.
The decidability of equivalence of DTAs is well-known
and can be done in polynomial time~\cite{DBLP:journals/actaC/GecsegS80,DBLP:books/others/tree1984}.
To obtain a total transducer we add for each pair $(q,f)$, $q \in Q$ and $f \in \Sigma$
that has no rule a new rule $q(f(\underline{x}),\underline{y}) \to \bot$,
where $\bot$ is an arbitrary symbol in $\DeltaO$ of rank zero.
\begin{example}
In Example \ref{ex:binary} we discussed how to adjust the transducer from the introduction
to our formal definition. We therefore had to introduce additional rules to obtain a
total transducer. Now we still add rules for the same pairs $(q,f)$
but only with right-hand sides $\bot$.
Therefore the original domain of the transducer is given by a DTA $D = (R,\Sigma,r_0,\delta_D)$
with the rules $r_0(g(x_1,x_2)) \to (r(x_1), r(x_2))$, $r(f(x_1,x_2)) \to (r(x_1),r(x_2))$ and
$r(i) \to (\ )$ for $i = 1,2,3$.
\qed
\end{example}

\begin{corollary}
The equivalence of deterministic separated basic MTTs with a \emph{partial} transition
function is decidable.
\end{corollary}

Next, we show that our result can be used to decide the equivalence of total separated
basic MTTs with look-ahead. A total macro tree transducer with
regular look-ahead ($\text{MTT}^{\text{R}}$)
is a tuple $(Q,\Sigma,\Delta,\delta,R,\delta_R)$ where $R$ is a finite set of look-ahead states
and $\delta_R$ is a total function from $R^k \to R$ for every $f \in \Sigma^{(k)}$.
Additionally we have a deterministic bottom-up tree automaton $(P,\Sigma,\delta,-)$
(without final states).
A rule of the MTT is of the form
$$q(f(t_1,\ldots, t_k),y_1,\ldots, y_k) \to t \hspace{1cm} \angl{p_1,\ldots, p_k}$$
and is applicable to an input tree $f(t_1,\ldots,t_k)$ if the look-ahead automaton
accepts $t_i$ in state $p_i$ for all $i = 1,\ldots, k$.
For every $q, f, p_1,\ldots, p_k$ there is exactly one such rule.
Let $N_1 = (Q_1,\Sigma_1,\Delta_1,\delta_1,R_1,\delta_{R1}), N_2 = (Q_2,\Sigma_2,\Delta_2,\delta_2,R_2,\delta_{R2})$
be two total separated basic MTTs with look-ahead. We construct
total separated basic MTTs $M_1, M_2$ \emph{without} look-ahead as follows.
The input alphabet contains for every $f \in \Sigma$ and $r_1,\ldots,r_k \in R_1$,
$r'_1,\ldots, r'_k \in R_2$
the symbols $\angl{f,r_1,\ldots,r_k,r'_1,\ldots,r'_k}$.
For $q(f(x_1,\ldots, x_k),\underline{y}) \to p[T_1,\ldots,T_m]\ \angl{r_1,\ldots, r_k}$
and $q'(f(x_1,\ldots, x_k),\underline{y'}) \to p'[T'_1,\ldots,T'_m]\ \angl{r'_1,\ldots, r'_k}$
we obtain for $M_1$ the rules 
$$\hat{q}(\angl{f(x_1,\ldots, x_k), r_1,\ldots, r_k, r'_1,\ldots, r'_k},\underline{y}) \to p[\hat{T}_1,\ldots, \hat{T}_m]$$
with $\hat{T}_i = \hat{q}_{i}(\angl{x_{j_i}, \hat{r_1},\ldots, \hat{r_l},\hat{r'_1},\ldots, \hat{r'_l}},Z_{i})$ if 
$T_i = q_i(x_{j_i},Z_i)$ and
$q_i(x_{j_i},\underline{y}) \to \hat{T_i}\ \ \angl{\hat{r_1},\ldots,\hat{r_l}}$ and
$q'_i(x_{j_i},\underline{y'}) \to \hat{T'_i}\ \ \angl{\hat{r'_1},\ldots,\hat{r'_l}}$.
If $T_i = y_{j_i}$ then $\hat{T}_i = y_{j_i}$.
The total separated basic MTT $M_2$ is constructed in the same lines.
Thus, $N_i$, $i=1,2$ can be simulated by $M_i$, $i=1,2$, respectively,
if the input is restricted to the regular tree language of new input trees that represent
correct runs of the look-ahead automata.

\begin{corollary}
The equivalence of total separated basic MTTs with regular look-ahead is decidable
in polynomial time.
\end{corollary}

Last, we consider separated basic MTTs that concatenate strings instead of trees in the parameters.
We abbreviate this class of transducers by $\text{MTT}^\text{yp}$.
Thus, the alphabet $\DeltaI$ is not longer a ranked alphabet but a unranked alphabet which
elements/letters can be concatenated to words. 
The procedure to decide equivalence of $\text{MTT}^\text{yp}$ is essentially the same as
we discussed in this paper but instead of conjunctions of equations of trees over 
$\DeltaI \cup Y$ we obtain conjunctions equations of words. 
Equations of words is a well studied problem ~\cite{Makanin1977,DBLP:conf/focs/Plandowski99,Lothaire2002}. In particular, the confirmed Ehrenfeucht conjecture states that each conjunction
of a set of word equations over a finite alphabet and using a finite number of variables, 
is equivalent to the conjunction of a finite subset of word equations \cite{DBLP:journals/tcs/Honkala00a}. 
Accordingly, by a similar argument as in section \ref{s:poly}, the sequences of 
conjunctions $\Psi_{b,q}^{(h)}(z),{\Psi'}_{b,q'}^{(h)}(z),\Phi_{b,(q,q')}^{(h)},h\geq 0$,
are ultimately stable. 

\begin{theorem}
The equivalence of total separated basic MTTs that concatenate words instead of trees in
the parameters ($\DeltaI$ is unranked) is decidable.
\end{theorem}

\section{Related Work}

\begin{figure}[htb]
\centering
\begin{tikzpicture}
  \node[] at (0,1em) (DMTT) {\normalsize \underline{DMTT}};
    \node[below =5mm of DMTT] (ATT) {\normalsize $\text{\underline{ATT}}$};    
    \node[below =5mm of ATT] (nATT) {\normalsize $\text{\underline{nATT}}$};
    \node[below=5mm of nATT] (snATT) {\normalsize $\text{snATT}$};
    \node[below right =5mm and 7mm of DMTT] (DMSOTT) {\normalsize $\text{DMSOTT}$};
    
    \node[below=5mm of DMSOTT] (lATT) {\normalsize $\text{lATT}$};
    
    \node[below left=5mm and 7mm of DMTT] (bDMTT) {\normalsize $\text{\underline{bDMTT}}$};
    \node[below=5mm of bDMTT] (sbDMTT) {\normalsize $\text{sbDMTT}$};
    \node[below=5mm of sbDMTT] (DT) {\normalsize $\text{DT}$};
    
    \draw (DMTT) -- (ATT);
    \draw (ATT) -- (nATT);
    \draw (nATT) -- (snATT);
    
    \draw (DMTT) -- (bDMTT);
    \draw (bDMTT) -- (sbDMTT);
    \draw (sbDMTT) -- (DT);
    
    \draw (ATT) -- (lATT);
    \draw (DMTT) -- (DMSOTT);
    \draw (DMSOTT) -- (lATT);

    \draw(bDMTT) -- (nATT);
    \draw(sbDMTT) -- (snATT);
    \draw(DT) -- (ATT);
\end{tikzpicture}
\caption{Classes with and without (underlined) known decidability of equivalence}\label{fig:classes}
\end{figure}
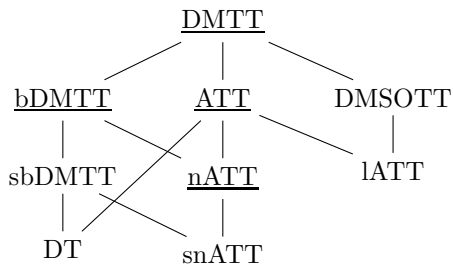

For several subclasses of attribute systems equivalence is known to be decidable.
For instance, attributed grammars without inherited attributes are equivalent
to deterministic top-down tree transducers (DT)~\cite{Engelfriet1980,DBLP:journals/tcs/CourcelleF82}. 
For this class equivalence was shown to be decidable by Esik~\cite{DBLP:journals/actaC/Esik81}. 
Later, a simplified algorithm was provided in~\cite{DBLP:journals/jcss/EngelfrietMS09}.
If the tree translation of an attribute grammar is of linear size increase, then equivalence is decidable,
because it is decidable for deterministic macro tree transducers (DMTT) of linear size increase. 
This follows from the fact that
the latter class coincides with the class of (deterministic) MSO definable tree translations 
(DMSOTT)~\cite{DBLP:journals/siamcomp/EngelfrietM03} for
which equivalence is decidable~\cite{DBLP:journals/ipl/EngelfrietM06}.
\ignore{
For deterministic MTTs that produce only monadic output trees equivalence is decidable.
This follows from the fact that such transducers are equivalent (by considering monadic trees
as strings)
to determinisitic top-down tree-to-string
transducers~\cite{DBLP:journals/acta/EngelfrietV88,DBLP:journals/iandc/EngelfrietM99} 
for which equivalence was recently proved to be decidable~\cite{DBLP:journals/jacm/SeidlMK18}.
For the inclusions of $\text{DT}^{\text{R}}$ in $\text{MTT}_{\text{mon}}^{\text{R}}$ 
and of
DMSOTT in $\text{MTT}_{\text{mon}}^{\text{R}}$, consider the output tree as 
a string (with brackets) and turn that string into a monadic tree.}
Figure~\ref{fig:classes} shows a Hasse diagram of classes of translations
realized by certain deterministic tree transducers.
The prefixes  ``l'', ``sn'', ``b'' and ``sb'' mean 
``linear size increase'', 
``separated non-nested'',
``basic'' and 
``separated basic'', respectively.
A minimal class where it is still open whether equivalence is decidable is the class of \emph{non-nested} attribute systems (nATT)
which, on the macro tree transducer side, is included in the class of \emph{basic} deterministic macro tree transducers (bDMTT).

For deterministic top-down tree transducers, equivalence can be decided in EXPSPACE, 
and in NLOGSPACE if the transducers are total~\cite{DBLP:journals/ijfcs/Maneth15}.
For the latter class of transducers, one can decide equivalence 
in polynomial time by transforming the transducer into
a canonical normal form
and then checking isomorphism of the 
resulting transducers~\cite{DBLP:journals/jcss/EngelfrietMS09}.
In terms of hardness, we know that equivalence of deterministic top-down tree
transducers is EXPTIME-hard. For linear size increase deterministic macro tree transducers
the precise complexity is not known (but is at least NP-hard).
More compexity results are known for other models of tree transducers such
as streaming tree transducers~\cite{DBLP:journals/jacm/AlurD17}, 
see~\cite{DBLP:journals/ijfcs/Maneth15} for a summary.

\section{Conclusion}

We have proved that the equivalence problem for separated non-nested 
attribute systems can be decided in polynomial time. 
In fact, we have shown a stronger statement, namely that in polynomial
time equivalence of \emph{separated basic total deterministic
macro tree transducers} can be decided.
To see that the latter is a strict superclass of the former, consider
the translation that takes a binary tree as input, and outputs the
same tree, but under each leaf a new monadic tree is output
which represents the inverse Dewey path of that node.
For instance, the tree $f(f(a,a),a)$ is translated into the tree
$f(f(a(1(1(e))),a(2(1(e)))),a(2(e)))$. A macro tree transducer of the
desired class can easily realize this translation using a rule of the form
$q(f(x_1,_2),y)\to f(q(x_1,1(y)),q(x_2,2(y)))$.
In contrast, no attribute system can realize this translation. The reason is
that for every attribute system, the number of distinct output subtrees is
linearly bounded by the size of the input tree. For the given translation there
is no linear such bound (it is bounded by $|s|\log(|s|)$). 

The idea of ``separated'' to use different output alphabets, is related to the
idea of transducers 
``with origin''~\cite{DBLP:conf/icalp/Bojanczyk14,DBLP:journals/iandc/FiliotMRT18}. 
In future work we would like to 
define adequate notions of origin for macro tree transducer, and prove that
equivalence of such (deterministic) transducers with origin is decidable.

\bibliographystyle{abbrv}
\bibliography{literature}
\appendix

\section{Proof of Lemma \ref{l:prefix}}

Let $M$ be a total separated basic MTT $M$, $D$ a given DTA.
Let $t\in\dom(\pi(q))$ be a smallest input tree of a state $q$ of $M$.
The $\DeltaO$-prefix of every state $q$ of $M$ relative to $D$
can be computed in $\mathcal{O}(|t|\cdot |M|)$.

\begin{proof}
The proof is similar to the one of \cite[Theorem 8]{DBLP:journals/jcss/EngelfrietMS09} 
for top-down tree transducers.
This construction can be carried over as, for the computation of $\DeltaO$-prefixes,
the precise contents of the output parameters $y_j$ can be ignored.

The $\DeltaO$-prefixes can be computed with the following system of in-equations
over the complete lattice $\mathcal{P}_{\DeltaO}$ with
unknowns $Y_{q}$, $q \in Q$:

For each rule $q(f(x_1,\ldots,x_k),y_1,\ldots,y_l) \to T$ such that
there is a tree $f(x_1,\ldots, x_k) \in \dom(\pi(q))$
we have the in-equation
\begin{eqnarray}
Y_{q} &\sqsupseteq& p[Z_1,\ldots,Z_m]	\label{e:prefix}
\end{eqnarray}
if $T = p[T_1,\ldots,T_m]$ for some pattern $p\in\T_{\DeltaO\cup\{\top\}}$
and terms $T_1,\ldots, T_m$ which are either variables $y_j$ or of the form
$q'(x_i, t'_1, \ldots, t'_l)$, $1 \leq i \leq k$
for some $q'\in Q$ where
\begin{eqnarray*}
Z_i &=& \left\{
	\begin{array}{ll}
	\top	&\text{if}\; T_i= y_j	\\
	Y_{q'}	&\text{if}\; T_i = q'(x_i, t'_1, \ldots, t'_{l})
	\end{array}\right.
	\label{e:prefix_cont}
\end{eqnarray*}
Since all right-hand sides of in-equations \eqref{e:prefix} are monotonic in their arguments, the system has 
a unique \emph{least} solution.
%
Moreover, we observe that the right-hand sides are distributive w.r.t.\ the least upper bound of patterns in the arguments.
I.e., for sets of patterns $S_{i}$ with $\bar{p_i} = \lub S_i$, 
$p[\bar{p}_1,\ldots, \bar{p}_m] = \lub \{p[p_1,\ldots,p_m] \mid p_i \in S_i\}$.

First, we show that the $\DeltaO$-prefixes $\prefO(q)$, $q\in Q$ build a solution for
the system of in-equations (\ref{e:prefix}).
Let $q(f(x_1,\ldots, x_k),y_1,\ldots,y_l) \rightarrow p[T_1,\ldots T_m]$ 
with $T_i = y_{j_i}$ or $T_i = q_i(x_{j_i},\underline{T_i})$
then
\begin{eqnarray*}
 \prefO(q) &=& \lub \{\sem{q}(t,\underline{T'}) \mid t\in\dom(\pi(q)), \underline{T'}\in\T_{\DeltaI}^l\} \\
  	    &\sqsupseteq& \lub \{\sem{q}(f(\underline{t}),\underline{T'}) \mid 
				f(\underline{t}) \in \dom(\pi(q)), \underline{T'}\in\T_{\DeltaI}^l\} \\
	    &=& \lub \{p[
 		\sem{T_1}\underline{t}\ \underline{T'}, \ldots, 
 	        \sem{T_m}\underline{t}\ \underline{T'}] 
		  \mid t_i\in\dom(\pi(q_i)), \underline{T'} \in\T_{\DeltaI}^l\}	\\
\end{eqnarray*}
By distributivity of pattern substitution, the latter pattern equals $p[\bar p_1,\ldots,\bar p_m]$ where
\begin{eqnarray*}
\bar p_i &=& \begin{cases}
                \top, & \text{whenever}\quad T_i = y_{j_i}\\
                \parbox[b]{4.9cm}{$\lub \{\sem{q_i}(t,\underline{T_{j_i}}[T'_1/y_1,\ldots,T'_l/y_l])\mid t\in\dom(r_i),T'_j\in\T_{\DeltaI}\}$,}& \text{if}\quad T_i = q_i(x_{j_i},\underline{T_{j_i}})\\
            \end{cases}\\
        &=& \prefO(q_i)
\end{eqnarray*}
\ignore{
\[
\begin{array}{llll}
\bar p_i &=& \top	&\text{, whenever}\quad T_i = y_{j_i}	\\
\bar p_i &=& \lub \{\sem{q_i}(t,\underline{T_{j_i}}[T'_1/y_1,\ldots,T'_l/y_l])\mid \\
&& \hspace{1cm} t\in\dom(\pi(q_i)),T'_j\in\T_{\DeltaI}\}
			& \text{, if}\quad T_i = q_i(x_{j_i},\underline{T_{j_i}})	\\
	&=&\prefO(q_i)
\end{array}
\]
}
Therefore, $\prefO(q),q\in Q$, satisfies all constraints.

Now consider \emph{any} solution $p'_{q}, q\in Q$ of the given constraint system. We claim that
then $\prefO(q)\sqsubseteq p'_{q}$ holds for all $q$. Accordingly, 
$\prefO(q)$ is the \emph{least} solution to the given constraint system.
In order to prove the claim, we proceed by induction on the input trees $t\in \dom(\pi(q))$.
Assume that $\pi(q)(f(x_1,\ldots, x_k)) \to (r_1(x_1),\ldots,r_k(x_k))$
and $q(f(t_1,\ldots, t_k),\underline{y}) \rightarrow p[T_1,\ldots, T_m]$
is the rule of the MTT for $q$ and $f$ where $T_i$ either equals some parameter $y_{j_i}$ or an
expression $q_i(x_{j_i},\underline{S_{i}})$.  
In that latter case, we have by
inductive hypothesis for the $t_i\in\dom(\pi(q_i))$, that $p'_{q_i}$ is a prefix of
$\sem{q_i}(t_i,\underline S)$ for \emph{every} tuple $\underline S\in\T_{\DeltaI}^l$,
i.e., $p'_{q_i}\sqsupseteq \sem{q_i}(t_i,\underline S)$ holds.
For each $i$ with $T_i = y_{j_i}$, define $p'_i$ as the pattern $\top$.
Then, $p[p'_1,\ldots,p'_m]$ is a prefix of 
$\sem{q}(t,\underline S)$, i.e., 
$p[p'_1,\ldots,p'_m]\sqsupseteq \sem{q}(t,\underline S)$ holds.
Since $p'_q$ is a solution of the system of in-equations, we also have that
$p'_{q} \sqsupseteq p[p'_1,\ldots,p'_m]$. Thus, the claim follows by transitivity of $\sqsupseteq$.

To compute the $\DeltaO$-prefixes for every $q \in Q$ we first compute some output tree
$t_q = q(t,\underline{Y})$ where $t$ is a minimal input tree in $\dom(\pi(q))$ and $\underline{Y}$ a minimal vector of terms over $\DeltaI$.
These $t_q$ can be computed in polynomial time and serve as lower bound for $\prefO(q)$, i.e., $t_q \sqsubseteq \prefO(q)$.
We therefore take $t_q$ as inital values for $Y_{q}$ in the fixpoint iteration of the constraint system.
Since in each iteration at least one subtree of the current value of $Y_{q}$ has to be replaced the fixpoint iteration
ends after a polynomial time of iterations.
\qed
\end{proof}

\section{Proof of Lemma \ref{l:earliest}}

For every pair $(M,A)$ consisting of a total deterministic separated basic MTT $M$ and axiom $A$
and a given DTA $D$, 
an equivalent pair $(M',A')$
can be constructed so that
$M'$ is a total deterministic separated basic MTT that is $D$-earliest.
Let $t$ be an output tree of $(M,A)$ for a smallest input tree $t \in\dom(\pi(q))$
where $q$ is the state occuring in $A$.
Then the construction runs in $\mathcal{O}(|t|\cdot|(M,A)|)$.

\begin{proof}

Let $M=(Q,\Sigma,\Delta,\delta)$ and $A$ be an axiom.
The new transducer $M'=(Q',\Sigma,\Delta,\delta')$ is defined as follows:
The set $Q'$ of states of $M'$ are pairs $\angl{q,v}$ where $q\in Q$ is a state of $M$,
and $v$ the position of a $\top$ leaf in the pattern $\prefO(q)$,
while the new axiom $A'$ and the new transition relation 
$\delta'$ are obtained as follows.
Let the axiom $A$ of $M$ be of the form 
$A=p[q_1(x_1,\underline{T_1}),\ldots,q_m(x_1,\underline{T_m})]$ with
$\underline{T_1},\ldots,\underline{T_m}$ vectors of output parameters and
for $j=1,\ldots,m$, let
$t_j = \prefO(q_j)[\angl{q_j,v_{j1}}(x_1,\underline{T_j}),\ldots,
			 \angl{q_j,v_{jr_j}}(x_1,\underline{T_j})]$.
Here $v_{j1},\ldots,v_{jr_j}$ is the sequence of positions of $\top$-leaves in $\prefO(q_j)$.
Then the new axiom is given by $A' = p[t_1,\ldots,t_m]$.
Now assume that $q(f(x_1,\ldots,x_k),\underline y)\to p[u_1,\ldots,u_m]$
where for $i=1,\ldots, m$, $u_i$ either equals a parameter $y_{j_i}$ or
a call $q_i(x_{j_i},\underline{T_i})$. 
Let $I$ denote the set of indices where the latter is the case. 
Let $\bar u_i$ denote the term $\prefO(q_i)$ if $i\in I$ and
$u_i$ otherwise.
By construction, $p[\bar u_1,\ldots,\bar u_m]$ is a prefix of $\prefO(q)$.
Let 
$t'_i = \prefO(q_i)[\angl{q_i,v_{i1}}(x_{j_i},\underline{T_i}),\ldots,\angl{q_i,v_{ir_i}}(x_{j_i},\underline{T_i})]$ if $i\in I$ 
($v_{i1},\ldots,v_{ir_i}$ is the sequence of positions of $\top$ leaves in $\prefO(q_j)$)
and $y_{j_i}$ otherwise. 
Then $\prefO(q)[s_1,\ldots,s_r] = p[t'_1,\ldots,t'_m]$ for some $s_1,\ldots,s_r$.
Assuming that $v_1,\ldots,v_r$ are the positions of $\top$-leaves in $\prefO(q)$,
we set 
\[
\angl{q,v_j}(f(x_1,\ldots,x_k),\underline y) \to s_j
\]
for $j=1,\ldots,r$. This ends the construction of the new transducer $M'$ with the corresponding axiom $A'$.
Note, that the mapping $\pi'$ from states of $M'$ to states of $D$ is given by
$\pi'(\angl{q,v}) = b$ if and only if $\pi(q) = b$ with
$\angl{q,v} \in Q', q \in Q, b \in B$.

From the construction follows that $M'$ is an earliest total deterministic separated basic MTT.
We show that $(M, A)$ is equivalent to $(M', A')$ by induction over the structure of the input trees.
The semantics of the pairs $(M,A)$ and $(M',A')$ is defined as 
$\sem{(M,A)}(t,\underline{T}) = p[\sem{q_1}(t,\underline{T_1}), \ldots, \sem{q_m}(t,\underline{T_m})]$ and
$\sem{(M',A')}(t,\underline{T})=p[\sem{t_1},\ldots,\sem{t_m}]$ with 
$\sem{t_j} = \prefO(q_j)[\sem{\angl{q_j,v_{j1}}}(t,\underline{T_j}),\ldots,
			      \sem{\angl{q_j,v_{jr_j}}}(t,\underline{T_j})]$
for $j = 1,\ldots,m$. Therefore, it suffices to show for every state $q\in Q$, every input tree $t\in\dom(\pi(q)) = \dom(\pi'(q'))$
and every vector of output parameters $\underline{T}\in \T_{\DeltaI}^l$, that
\[
\sem{q}(t, \underline{T}) = \prefO(q)[\sem{\angl{q,v_{1}}}(t, \underline{T}),\ldots,
						  \sem{\angl{q,v_{r}}}(t, \underline{T})]
\]
holds if the subtrees $\top$ of $\prefO(q)$ occur at positions $v_1,\ldots,v_r$, respectively.


For $j=1,\ldots,m$, there are thus nodes $v_{j1},\ldots,v_{jr_j}$ in $\prefO(q_j)$ 
such that for each input tree $t\in\dom(\pi(q_j))$ and vector of output parameters $\underline{T}\in\T_{\DeltaI}^l$,
$\sem{q_j}(t,\underline{T}) = \prefO(q_j)[\sem{\angl{q_j,v_{j1}}}(t,\underline{T}),\ldots,
						\sem{\angl{q_j,v_{jr_j}}}(t,\underline{T})]$. 
Note, that in the base case when $t=g$ and $g\in \Sigma$ has rank $0$ ($m=1$), the right-hand side of a rule of $q$
is a term $t\in \T_{\DeltaO\cup Y}$, i.e., $q(t,\underline{y})\rightarrow p[y_{i1},\ldots, y_{ik}]$.
Then for $i=1,\ldots,r$, $\angl{q,v_i}(t,\underline{y}) \rightarrow s_i$ where $\prefO(q)[s_q,\ldots,s_r] = p[y_{i1},\ldots,y_{im}]$.
Therefore,
\[
\begin{array}{lll}
\sem{M,A}(t,\underline{T})	&=&
p[\sem{q_1}(t,\underline{T}),\ldots,
  \sem{q_m}(t,\underline{T_m})]	\\
&=&p[
	\begin{array}[t]{l}
\prefO(q_1)[
	\sem{\angl{q_1,v_{11}}}(t,\underline{T_1}),\ldots,
	\sem{\angl{q_1,v_{1r_1}}}(t,\underline{T_1})],\ldots,	\\
\prefO(q_m)[
	\sem{\angl{q_m,v_{m1}}}(t,\underline{T_m}),\ldots,
	\sem{\angl{q_m,v_{mr_m}}}(t,\underline{T_m})]\;]
	\end{array}		\\
	&=& \sem{M',A'}(t,\underline{T})
\end{array}
\]
with $t \in \dom(\pi(q)) = \dom(\pi'(q'))$.
Accordingly, for all input trees $t \in \dom(\pi(q))$ and parameter vectors $\underline{T} \in \T_{\DeltaI}^l$, $\sem{M,A}(t,\underline{T}) = \sem{M',A'}(t,\underline{T})$.

Assume that $n$ is the number of states of $M$, and $D$ is the maximal size of a common $\DeltaO$-prefix.
Thus $D$ is at most of the size of a right-hand side of a rule for a symbol of rank $0$.
Then each state $q$ of $M$ gives rise to at most $n\cdot D$ states $\angl{q,v}$ of $M'$ where
each right-hand side of $\angl{q,v}$ is at most $D+1$ times the size of the corresponding right-hand side of $q$.
Thus, the size of $M'$ is polynomial in the size of $M$. Likewise, the axiom $A'$ is only a factor of $D+1$ larger than
the axiom $A$. 
Since the $\DeltaO$-prefixes of states of $M$ can be computed
in polynomial time, the overall algorithm runs in polynomial time.
\qed
\end{proof}

\end{document}